# Nonlinear analysis of EEG complexity in episode and remission phase of recurrent depression

Cukic Milena PhD[1, *], Stokic Miodrag PhD[2,3], Radenkovic Slavoljub[4], Ljubisavljevic Milos MD PhD[5], Simic Slobodan MD[6], Danka Savic[7] PhD

[1]Department of General Physiology and Biophysics, School of Biology, University of Belgrade, Serbia

[2]Life Activities Advancement Center, Belgrade, Serbia

[3]Institute for Experimental Phonetics and Speech Pathology, Belgrade, Serbia

[4]TomTom, Amsterdam, Netherlands

[5]Department of Physiology, College of Medicine and Health Sciences, UAE University, Al Ain, UAE

[6]Institute for Mental Health, Belgrade, Serbia

[7] Vinca Institute, Laboratory of Theoretical and Condensed Matter Physics 020/2, University of Belgrade, Belgrade, Serbia

**Author note**

Authors of this paper are Milena Čukić, PhD, Studentski trg 16, 11 000 Belgrade, Serbia, tel. +381 60 0287704, e-mail: milena.cukic@gmail.com ; Miodrag Stokić, PhD, Gospodar Jovanova 35, 11 000 Belgrade, Serbia, tel. +381 11 3208 532, e-mail: m.stokic@iefpg.org.rs , Slavoljub Radenković, Oosterdoksstraat 114, 1011 DK Amsterdam, the Netherlands, tel. +381621775149, e-mail: sradenkovic@gmail.com; Slobodan Simić, M.D, Palmotićeva 37, Belgrade, Serbia, tel.+381 11 330 7543, e-mail:slobodansimic5@gmail.com ; Miloš

Ljubisavljević, M.D,PhD, P. O. Box 17666, Al-Ain, United Arab Emirates, tel. +971 3 7137 707, e-mail: milos@uaeu.ac.ae ; Danka Savić, PhD, Mike Petrovica Alasa 12-14, 11001 Belgrade, Serbia, danka.s@sbb.rs.

Corresponding author has been authorized by all other authors to act on their behalf in all matters pertaining the publication of the manuscript. Order of the names has been agreed by all authors.

* Correspondence concerning this article should be addressed to Milena Čukić, Ph.D., Studentskitrg 16, 11 000 Belgrade, Serbia, tel. +381 11 3208 532 (Current address: KoninginWilhelminaplein 644, 1062 KS Amsterdam, The Netherlands, +31615178926), e-mail: milena.cukic@gmail.com

# Abstract

Background: Biomarkers of Major Depressive Disorder(MDD), its phases and forms have long been sought. Research indicates that the complexity measures of the cortical electrical activity (EEG) might be candidates for this role.

Aims: To examine whether the complexity of EEG activity, measured by Higuchi's fractal dimension (HFD) and sample entropy (SampEn), differs between healthy subjects, patients in remission and episode phase of the recurrent depression and whether the changes are differentially distributed between hemispheres and cortical regions.

Methods: Resting state EEG with eyes closed was recorded from 26 patients suffering from recurrent depression and 20 age and sex-matched healthy control subjects. Artefact-free EEG epochs were analyzed by in-house developed programs running HFD and SampEn algorithms.

Results: Depressed patients had higher HFD and SampEn complexity compared to healthy subjects. Surprisingly, the complexity was even higher in patients who were in remission than in those in the episode. Altered complexity was present in the frontal and centro-parietal regions when compared to control group. The complexity in frontal and parietal regions differed between the two phases of depressive disorder. SampEn manifested higher sensitivity than HFD in some cortical areas.

Conclusions: Complexity measures of EEG distinguish between the three groups. Further studies are needed to establish whether these measures carry a potential to aid clinically relevant decisions about depression.

**Introduction**

Major Depressive Disorder (MDD) is a serious mental illness associated with protracted personal suffering, and significant social and functional impairment. It has become the leading cause of ill health and disability worldwide, before heart diseases, arthritis and many forms of cancer. Depression has a strong tendency to reoccur - a significant number of patients will suffer from at least one more episode after the first one, reaching four episodes on average during the lifetime [1]. In these patients, the risk of new episodes rises significantly with each subsequent recurrence although the course of the disease can be unique [1]. Hence, the decision to stop the therapy, or to initiate the maintenance therapy to prevent a relapse in patients with recurrent depression who have achieved remission, often presents a significant clinical challenge [2]. Therefore, finding accurate and reliable biomarkers that can help differentiate MDD episode from remission is of considerable importance. Also, there is a need to distinguish the (onset) episode of bipolar disorder from a depressive episode in Major Depression [3] in order to treat it appropriately.

Stress hormones, most of all cortisol, have long been the main candidate biomarkers for depression, but the results of numerous studies are inconclusive [4]. It seems that combined biomarkers are more promising than any single one [5], but there still are no biomarkers that are sufficiently sensitive and specific [6].

Methods based on the theory of nonlinear dynamics are justly included in the quest for biomarkers. They provide more accurate information than classical spectral analysis and have already enabled insights into neural activity and connectivity in both healthy physiological processes and pathological conditions. The first study that used nonlinear measures in depression showed that EEG dynamics, measured by correlation dimension, in patients with Major

Depressive Disorder (MDD) is more predictable, i.e., less complex than in healthy subjects [7]. Contrary to this result, Lempel–Ziv complexity measure of the EEG signal is higher in depressed patients compared to control subjects, particularly in anterior brain regions [8]. Furthermore, patients suffering multiple depressive episodes do not recover dynamics found in healthy subjects as patients who had one depressive episode [9]. The analysis of EEG signal with Approximate Entropy also showed higher values in normal controls compared to depression patients [10]. Other studies utilizing Lempel-Ziv complexity and other complexity measures found either no difference between MDD and control subjects [11], or increased EEG complexity in patients with depression [8,11–13]. Analyses of EEG complexity with Higuchi's fractal dimension (HFD) also showed increased EEG complexity in patients with depression, particularly in the beta and gamma sub-bands mostly in the frontal area [11,12]. Application of wavelet-chaos methodology found significantly increased complexity in both parietal and frontal regions in MDD patients, both in full-band activity and in beta and gamma EEG sub-bands [13]. De la Torre-Lugue and Xavier Bornas [14] in their review study concluded that 'EEG dynamics for depressive patients appear more complex but may be more random than the dynamics of healthy non-depressed individuals'.

In this study, we explore the use of different measures of complexity of brain activity, estimated from the resting state electroencephalogram (EEG) records, in discriminating episode from remission phase in patients with recurrent depression, as well as from healthy subjects.

## Methods

*Subjects*

EEG data were recorded at the Institute for Mental Health in Belgrade, Serbia. The participants were 26 patients (17 women and 9 men) suffering from recurrent depression, 25 to 68 years old (mean 32.40, SD 10.16). As a control, we used EEG recordings of 20 age-matched (mean 30.14, SD 8.94) healthy controls (10 males, 10 females) with no history of any neurological or psychiatric disorders, recorded at the Institute for Experimental Phonetic and Speech Pathology in Belgrade, Serbia. All participants were right-handed, according to Edinburgh Handedness Inventory. The participants were informed about the experimental protocol and signed informed consent. The protocol was approved by the Ethical Committees of the participating institutions. All participants with depression were on medications and under the supervision of an experienced specialist in clinical psychiatry. Their diagnoses was made according to ICD-10 scale. The study compared three groups: healthy controls (C), and depressed patients (D) in an episode (E), and in remission (R).

*Data acquisition*

EEG was recorded in the resting state with 10/20 International system for electrode placement, using NicoletOne Digital EEG Amplifier (VIAYSYS Healthcare Inc. NeuroCare Group), with closed eyes without any stimulus (resting state EEG). EEGs were recorded from 19 electrodes in a monopolar montage (Electro-cap International Inc. Eaton, OH USA). The sampling rate was 1 kHz. The resistance was kept less than five KOhm. A bandpass filter was 0.5-70 Hz. The same setup was used for the control group, using Nihon Kohden Inc. apparatus. Since the protocol of recordings was the same, the fact that we used recordings from two different EEG producers' apparata did not introduce the difference between the groups [15].

Each recording lasted for three minutes. Participants were instructed to reduce any movement, staying in a comfortable sitting position with eyes closed. The EEG records of four subjects were discarded from further analyses because of the high level of muscle activity or blinking artifacts. Further, we used records from 22 patients and 20 healthy controls for this study. Half of the patients were recorded while they were in an acute episode, while the other half were in remission phase of the disease. Artifacts were carefully inspected and removed manually from the records by two independent experts. From each artifact-free record, we chose three epochs: one from the beginning, one from the middle and the last one close to the end of the record. Each epoch comprises of 5000 samples. Positions of epochs in each person's EEG recording were specified by the ordinal number of the first sample in that epoch. Therefore, for each subject, epoch, and electrode, two nonlinear measures were calculated.

*Data analysis*

Initially, the classical spectral analysis was performed by constructing spectral power maps (EEGLAB program [16]). Thereon, for all EEG epochs, Higuchi's fractal dimension (HFD) and Sample entropy (SampEn) were calculated. Fractal and SampEn maps were constructed on the whole spectrum, not dividing the signal into bands. The analysis was adopted as it has been shown that the Fourier analysis is redundant to fractal analyses [17]. The fractal dimension of EEG was calculated by using Higuchi's algorithm [18], demonstrated to be the most appropriate for electrophysiological data [19]. This method provides a reasonable estimate of the fractal dimension even if short signal segments are analyzed and it is computationally fast. HFD was also chosen because it is widespread in the EEG literature facilitating comparison of the results. We performed the Higuchi's algorithm [18], with the maximal length of an epoch $k_{max}= 8$, shown to perform the best for this type of signals [20]. HFD of a time series is a measure of its complexity and self-similarity in the time domain. HFD is not an integer, and the value of fractal dimension (FD) of waveforms (e.g. EEG) can range between 1 and 2. Higher self-similarity and complexity results in higher HFD [21]. Sample Entropy (SampEn) was computed according to the procedure by Richman and Moorman [22]. SampEn estimates the signal complexity by computing the conditional probability that two sequences of a given length *n*, similar for *m* points, remain similar within tolerance *r* at the next data point (when self-matches are not included). SampEn measures the irregularity of the data (the higher the values, the less regular signal) that is related to signal complexity [23]. SampEn was calculated using tolerance level of $r = 0.15$ times the standard deviation of the time series and $m = 2$, shown to be optimal for EEG [24]. Both HFD and SampEn were calculated for each electrode for the duration of signal (the epochs of artifact-free recorded EEG; three epochs from each recording), using the in-house written algorithm in Java programming language. It should also be noted that correlations with any medical data were not explored since the main aim of the study was to find independent nonlinear markers based on analysis of the EEG signal, which could be utilized as an additional tool in clinical practice.

*Statistical analysis*

Both HFD and SampEn values were used as an ensemble for analysis of variance (ANOVA with post-hoc Bonferroni correction, SPSS Statistics version 20.0, SPSS Inc, USA). The Kolmogorov Smirnov test showed that HFD and SampEn data were not normally distributed. To obtain data with normal distribution and to include it in the analysis of the difference in complexity in resting state EEG data, normalized values of SampEn and HFD obtained from epochs of recorded EEG were calculated as log10 normalization in SPSS. Normalized SampEn and HFD data were compared using ANOVA with factors state (Controls vs. Depression, and Controls vs. Episode vs. Remissions) and position of the electrode (1 to 19). For every electrode, ANOVA was repeated for each measure independently. Bonferroni correction was used where appropriate. For all analyses, probability values $p = 0.05$ were considered as statistically significant.

*Principal Component analysis (PCA)*

To reduce the dimensionality of the problem and decorrelate the measures (HFD and SampEn calculated from the same epochs extracted from the raw EEG signal), we utilized PCA [25] in order to obtain three principal components (PCs) corresponding to largest eigenvalues of the sample covariance matrix. We defined percentage of the explained variance by first three PCs as ratio between sums of variances of three PCs and original variables. Here we wanted to demonstrate the possibility of classification of previously calculated nonlinear measures, by utilizing only the first three components in order to see whether the data were separable. We used Matlab 15b for this calculation (MathWorks, Masacushets, USA).

## Results

*Spectral power maps*

The first level of analysis was to compare spectral power maps of low alpha (8-10Hz), high alpha (10-12Hz), and beta (13-30Hz) bands between healthy controls (C) and patients in a different phase of the disease (i.e. episode (E) or remission (R), Figure 1). Spectral power maps in low alpha band showed an overall decrease in both E and R groups in posterior regions (maximum at C3, C4, Cz, P3, P4, Pz, and T3) when compared to C group. In E and R group there was an increase in low alpha spectral power in the right prefrontal region (Fp2) and lateral right frontal region (F8) when compared to C group. There was a statistically significant effect of group (C, E, R) on low alpha (8-10 Hz) spectral power at following regions: *C3* - $F(2,40)=8.748$ p=0.021; *C4* - $F(2,40)=5.168$, p=0.04; *Cz* - $F(2,40)=6.381$, p=0.03; *P3* - $F(2,40)=5.748$, p=0.021; *P4* - $F(2,40)=4.286$, p=0.05; and *Pz* - $F(2,40)=13.939$, $p<0.001$. Spectral power maps in high alpha band showed a decrease of high alpha spectral power in right prefrontal (Fp2), left temporal (T3), and central (C3, Cz) regions in E and R groups when compared to C group. Also, there was a statistically significant effect of group (C, E, R) on high alpha (10-12Hz) spectral power at following regions: ***Fp2*** - $F(2,40)=9.211$, $p<0.01$; ***C3*** - $F(2,40)=4.096$, $p=0.05$; ***Cz*** - $F(2,40)=10.734$, $p<0.01$; and ***T3*** - $F(2,40)=8.633$, $p=0.012$.

Spectral power maps in beta band showed a decrease of beta (13-30 Hz) spectral power at frontal (Fz, Fp1, Fp2, F3, F4, F7, F8), and central-temporal (C3, T3) regions in E and R groups when compared to C group. In contrast, there was an increase of beta spectral power in posterior regions (P3, P4, Pz, T5, T6, O1, O2) in E and R groups when compared to C group. In the E group only there was a frontal (F3 > F4) and temporal-occipital (T5/O1 < T6/O2) asymmetry. Significant effect of group (C, E, R) was found at following cortical regions: ***Fp1*** - $F(2,40)=7.207$, $p=0.018$; ***Fp2*** - $F(2,40)=8.107$, $p<0.01$; ***F3*** - $F(2,40)=12.342$, $p<0.01$; ***F4*** - $F(2,40)=18.674$, $p<0.01$; ***Fz*** - $F(2,40)=7.997$, $p=0.023$; ***F7*** - $F(2,40)=4.922$, $p=0.04$; ***F8*** - $F(2,40)=6.817$, $p=0.033$; ***P3*** - $F(2,40)=7.002$, $p=0.018$; ***P4*** - $F(2,40)=6.699$, $p=0.03$; ***Pz*** - $F(2,40)=14.361$, $p<0.01$; ***T5*** - $F(2,40)=13.207$, $p<0.01$; ***T6*** - $F(2,40)=14.338$, $p<0.01$; ***O1*** - $F(2,40)=16.189$, $p<0.01$; and ***O2*** - $F(2,40)=16.917$, $p<0.01$.

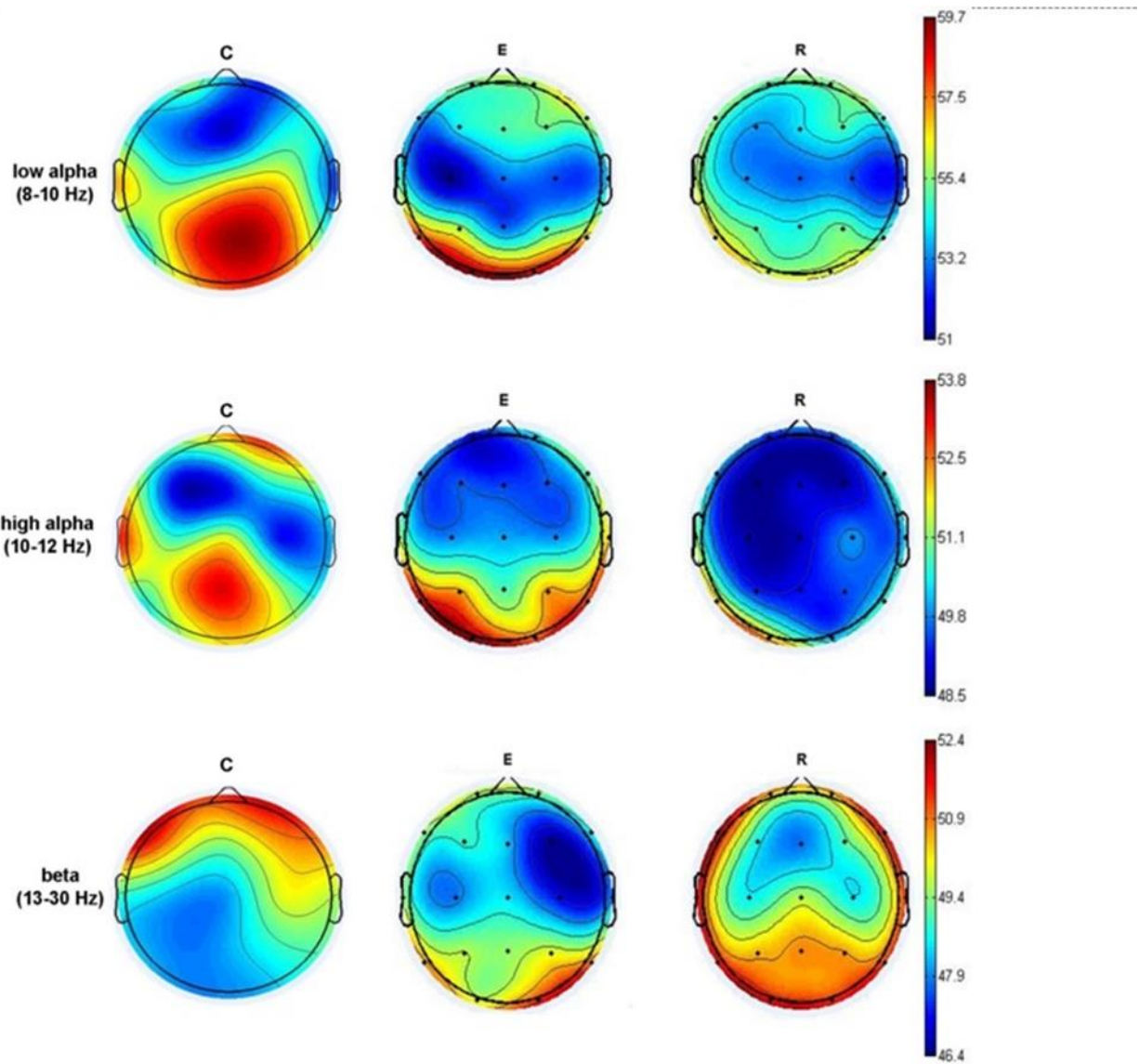

**Figure 1**. Spectral power maps are showing the difference between healthy persons (C) and those in the episode (E) and remission (R).

### *Higuchi's fractal dimension (HFD) and Sample entropy (SampEn)*

Figure 2 summarizes the difference in HFD and SampEn values in all three groups averaged across all electrodes. HFD values for healthy controls ranged from 1.0093 to 1.0530 (mean 1.0290), for patients in the episode from 1.01 to 1.3546 (mean 1.0876), and for those in remission from 1.0133 to 1.483 (mean 1.1136). SampEn values for healthy controls were 0.0764

to 0.3959 (mean 0.1613), in the episode group 0.2006 to 0.7866 (mean 0.3992) and in remission group 0.1983 to 0.861 (mean 0.4777).

ANOVA showed a significant effect of group on HFD ($F(2,2334)=30.831$, $p<0.01$) and on SampEn values ($F(2,2334)=38.863$, $p<0.01$). The post-hoc test showed that both HFD and SampEn values were lowest in the C group, followed by participants in the E group. Participants in the R group had the highest HFD and SampEn values ($HFD_{Control}<HFD_{Episode}<HFD_{Remission}$, $p<0.01$ for each comparison; $SampEn_{Control}<SampEn_{Episode}<SampEn_{Remission}$, $p<0.01$ for each comparison).

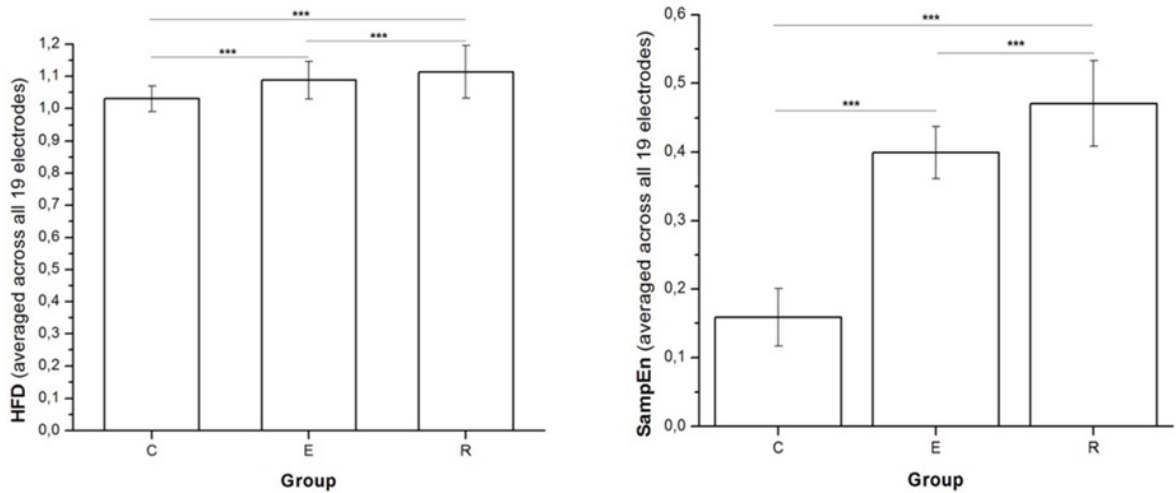

**Figure 2**. Averaged values for HFD (left) and SampEn (right) values averaged across all 19 electrodes, for all three groups (C- control, E-episode, and R-remission). Note the difference in scale for HFD and SampEn. *** $p<0.01$.

The next level of analysis was to determine effect of a Group (C, E, R) on HFD and SampEn values averaged across Regions of Interest – ROIs (ROI 1: left frontal regions Fp1, F3, F7; ROI 2: right frontal regions Fp2, F4, F8; ROI 3: Vertex regions Fz, Cz, Pz; ROI 4: left temporal-parietal-central-occipital regions T3, T5, P3, C3, O1; and ROI 5: right temporal-parietal-central-occipital regions T4, T6, P4, C4, O3). The data for ROI and group are shown in Figure 3. ANOVA showed a significant effect of group on HFD values for the ROI 1: F(2,366)=83.021, p<0.01, ROI 2: F(2,366)=85.400, p<0.01, ROI 3: F(2,366)=99.117, p<0.01, ROI 4: F(2,612)=78.908, p<0.01, and ROI 5: F(2,612)=94.547, p<0.01. The post-hoc test showed for each ROI that HFD is lowest in C group followed by E group while R group had the highest HFD values. A significant effect of group on SampEn values (ANOVA) was found for the ROI 1: F(2,366)=333.302, p<0.01, ROI 2: F(2,366)=308.381, p<0.01, ROI 3: F(2,366)=446.666, p<0.01, ROI 4: F(2,612)=464.020, p<0.01, and ROI 5: F(2,612)=492.307, p<0.01. The post-hoc test showed for each ROI that SampEn is lowest in C group followed by E group while R group had the highest SampEn values.

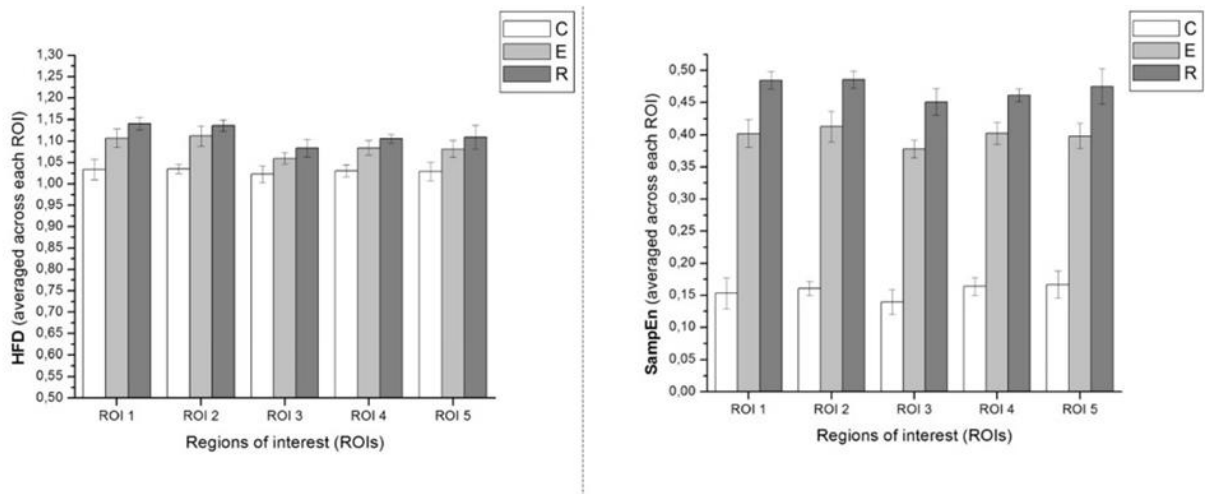

**Figure 3.** The comparison of HFD (left) and SampEn (right) values averaged across each Region of Interest (ROI). ROI1 included left frontal electrodes (Fp1, Fp3, F7). ROI2 included right frontal electrodes (Fp2, F4, F8). ROI3 included vertex electrodes (Fz, Cz, Pz). ROI4 included left temporal-parietal-central-occipital regions electrodes (T3, T5, P3, C3, O1) and ROI5 included right temporal-parietal-central-occipital electrodes (T4, T6, P4, C4, O3). C-control, E-episode, R-remission. Note the difference in scale for HFD and SampEn.

The final level of analysis was to determine the possible effect of a group on HFD and SampEn values for each electrode. ANOVA showed a significant effect of Group (C, E, R) on HFD values for each electrode (p<0.01). However, post-hoc test showed that significant difference between each group was found for electrodes Fp1, Fp2, Fz, P3, P4, T5, T6, and Cz (p<0.01). Post-hoc Bonferroni correction showed that for each electrode $HFD_{Control} < HFD_{Episode} < HFD_{Remission}$, p<0.05 for each comparison. A similar result was found for SampEn values. ANOVA showed a significant effect of Group (C, E, R) on SampEn values for each electrode (p<0.01). Post-hoc showed that this difference is driven by significantly lower SampEn values in C group when compared to E and R group. There was no significant difference between E and R group for each electrode. However, the post-hoc test showed that significant difference between each group was found for electrodes Fp1, Fp2, F4, F7, Fz, P3, P4, O2, T5, T6, Cz, and Pz (p<0.01) and that for each electrode $SampEn_{Control} < SampEn_{Episode} < SampEn_{Remission}$, p<0.05 for each comparison.

Two-way ANOVA found a statistically significant effect of factors Group (C, E, R) and ROI (frontal left, frontal right, vertex, TCPO left, TCPO right), as well as the interaction (Group x ROI) on HFD values (p < .001). The same analysis found a statistically significant effect of factors Group (p < .001) and ROI (p < .01) on SampEn values. No significant effect of the interaction was found.

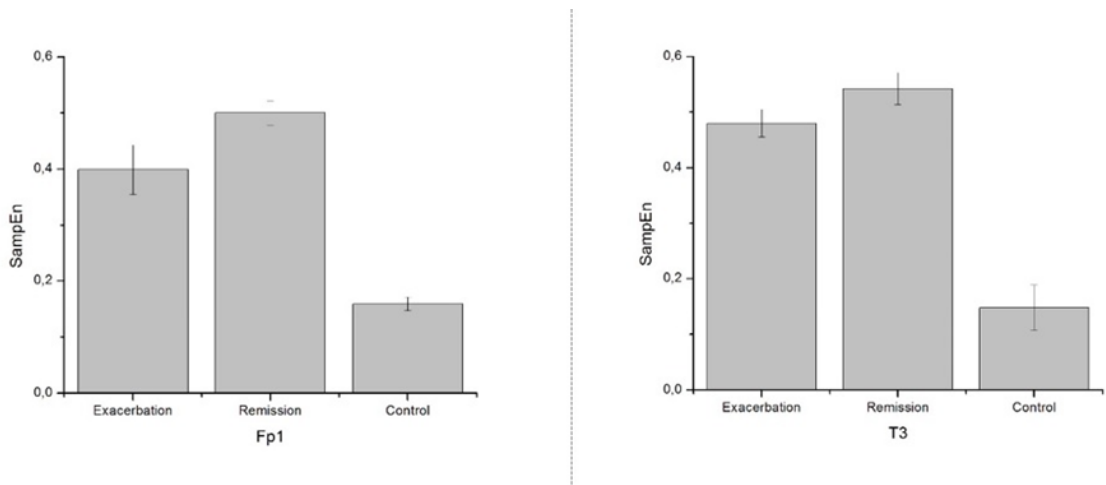

**Figure 4**: Results of (ANOVA) comparison of calculated sample entropy for certain electrodes which are particularly well discriminative: left, the values of SampEn for position Fp1 (fronto-parietal), and right, the values of calculated SampEn for EEG recorded from position T3 (temporal). The significant difference between those with depression and controls is particularly pronounced, and the difference between Exacerbation and Remission is also significant.

Figure 4 shows a spatial representation of significant differences of values of SampEn and HFD according to electrode position, i.e. *Fractal and SampEn maps*. The results show that SampEn is more discriminative regarding the number of electrodes with the statistically significant difference between C, E, and R.

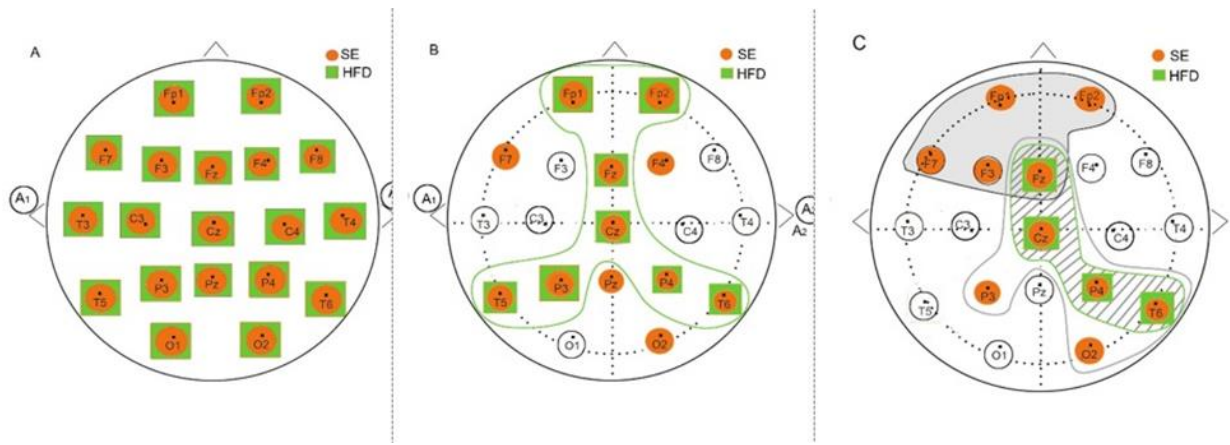

**Figure 5**: The spatial representation of significant differences in HFD and SampEn. In orange, electrodes from which SampEn values showed a significant difference in comparison to controls, and in green electrodes where HFD had a significant difference. The left panel show the difference between the whole group of patients compared to controls. Note that SampEn shows better discrimination than HFD. The middle panel shows the comparison of al three groups (C vs. E vs. R). A significant difference was found on 12 electrodes for SampEn, and on nine electrodes for HFD. Right panel shows the comparison of E and R groups. SampEn values were significantly different on Fp1, Fp2, F7, F3, P3, P4, T6, O2, Fz and Cz electrodes. HFD values were significantly different on Fz, Cz, P4, and T6 electrodes.

Figure 5-A shows that for both measures there is a significant difference between the patient and control group for the majority of electrodes. However, SampEn showed to discriminate R and E group (Figure 5-C) better. SampEn differed between R and E groups on Fp1, Fp2, P3, P4, O2, F7, T6, Fz, Cz electrodes, while HFD was different on Fz, Cz, P4, T6

electrodes only. When all three phases are examined (Figure 5-B) SampEn shows a significant difference on 12, and HFD on 8, out of 19 electrodes.

*Principal Component Analysis (PCA)*

With only the first three Principal Components (PCs) we want here to illustrate that those calculated values are separable. Again, SampEn gave more clear separation of the data when compared to HFD (on Figure 6, SampEn results are on the left and HFD on the right picture). In this study we did not intend to deal with further classification, although a high accuracy could be obtained for several machine learning algorithms.

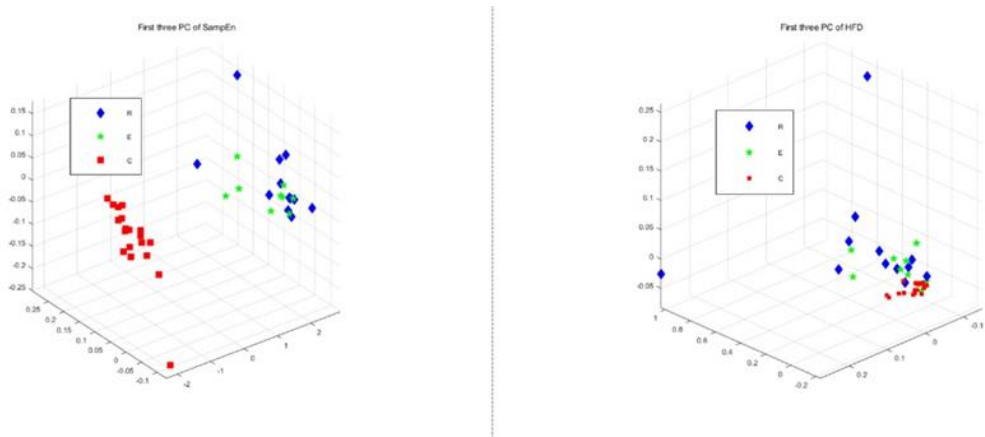

**Figure 6**: Principal component analysis was used to show the separability of data. We used only first three principal components. Blue diamonds are symbols for those who are diagnosed with depression and are in remission; green stars are depressed patients who are in exacerbation and red squares are representing control group. Panel left represents how separable are the values of SampEn calculated from EEG of subjects from these three groups; panel on the right represents the values of calculated Higuchi's fractal dimension which are obviously less separable data, but still distinctive.

## Discussion

The results show that both Higuchi fractal dimension (HFD) and SampEn nonlinear measures of resting state EEG signal discriminate between healthy controls and depressed patients, the latter having higher complexity. These differences are widely distributed and include frontal, midline (vertex), and temporal-parietal-occipital regions. Furthermore, the complexity differs significantly between episode and remission, being higher in remission than in the episode phase of the disease.

Although this last finding is counterintuitive (it would be expected that the remission state is closer to the healthy one in every respect), it might be in line with the observation of Willner et al. [26]: "It is evident that antidepressants do not normalize brain activity: mood and behavior are restored to normal, but the antidepressant-treated brain is in a different state from the non-depressed brain." According to the research on the first-episode depression, such brain is initially different from the non-depressed one and the differences are structural [27], as well as functional [28,29]. In terms of nonlinear dynamic systems, initial conditions of the depressed brain system are different from the non-depressed. Even if the dynamics of information processing were the same, these different initial conditions would direct the system to steady states distinctive from normal. However, the dynamics is probably also different due to compensatory mechanisms. For example, functional neuroimaging revealed reduced integrity of the uncinate fasciculus and enhanced functional connectivity of anterior cingulate cortex and medial temporal lobe in MDD [30]. The higher severity of depression, the more pronounced this negative structure-function relation. The authors suggest that the increased functional connectivity is a compensatory mechanism for decreased uncinate fasciculus integrity. Willner et al. came to a similar conclusion that decreased hippocampal functioning in depression causes an increase in the activity of the ventral "affective" system [26]. It is then easy to suppose that the enrichment of fronto-limbic connectivity and reorganization of circuits is accompanied by increment in complexity.

The EEG hallmark of depression is the presence of stable hemispheric asymmetry in the alpha spectral band, although the differences in other spectral bands were also demonstrated [31]. Interestingly, de Vinne et al [32] showed that frontal alpha asymmetry cannot be used as biomarker. At present, there is no consensus about the direction of change. Our spectral analysis shows that the power was decreased in alpha and high-alpha bands in majority of the cortical regions, but increased in beta bands in posterior regions in patients. This may point towards the presence of hyperactivity in posterior regions (alpha desynchronization) of the right hemisphere, which is known to process the negative emotional content [33].

As said in the Introduction, the results concerning complexity of EEG in depressed patients are discrepant. What may account for these opposing findings? One important aspect pertains to methodological differences between the studies, related to signal acquisition (number of EEG electrodes used, sampling frequency, pre-processing of raw signal, i.e., decomposition on bands and filtering), as well as experimental design (probing the emotional content, using different stimuli, performing cognitive task, etc.). The eyes-closed condition, unlike the eyes-open condition, allows measurement of the resting state arousal without the influence of cortical processing of the visual input in other bands on the complexity of brain dynamics. Also, it should be noted that we did not divide the spectrum of the signal to standard bands, but observed the changes in broadband. This is important as it has been shown that signal decomposition like Fourier, Wavelet, or cosine transformation can impact the result of a subsequent nonlinear analysis yielding erroneous results [34,35]. Other reasons may relate to inherent differences between nonlinear algorithms that are based on different theoretical frameworks [36,37]. Our results are in line with studies that also used Higuchi's fractal dimension [11,12], that was shown to be more accurate than Katz's algorithm [12]. The difference in complexity values between depressed and healthy subjects in our study were much larger than those reported in Bachman et al. [11]. Another possible source of difference is choosing different values for k in Higuchi's FD algorithm; Bachmann et al. [11] used 50 for their k value.

The results of Fractal and SampEn maps are in line with previous electrophysiological studies demonstrating the presence of stable frontal asymmetry [38,39] in MDD. However, in this study the signal was not divided to standard bands, hindering conclusion that current findings are directly related to the alpha band asymmetry. The results point to elevated complexity in frontal, central and right parieto-temporal regions. This is also in line with earlier EEG studies [40], which reported similar topographical changes in distribution.

It should be noted that we used HFD and SampEn, two nonlinear measures able to detect differential aspects of the signal under analysis. While HFD examines the complexity in the time domain, SampEn can characterize the irregularity of a signal or its predictability. They both showed higher complexity in patients with depression when compared to healthy control subjects. The difference was much more pronounced when examined by SampEn suggesting increased variability, or "irregularity" or unpredictability of the signal.

## Conclusions

The idea of EEG-based classification of depression is not entirely novel. Our study confirmed that it is possible to quantify the difference between depressed patients and controls by employing two complexity measures - HFD and SampEn, on resting state EEG. Furthermore, it showed, for the first time, that both measures could detect a statistically significant difference between depressed patients who were in episode and remission. Whether these and other non-linear measures may be used as potential clinical markers of disease stage or of the effectiveness of various treatments in MDD remains to be confirmed on larger groups of patients.

*Acknowledgments*

We want to thank Goran Petrovic for the input on the graphical representation of SampEn and HFD maps. This research was supported by the Ministry of Education, Science and Technological Development of the Republic of Serbia under Projects 178027 and 32032, and CMHS grant 31M201.

**Author Contributions:** M. C. - designed the study, analyzed the data and drafted the manuscript; M. S. - recorded EEG data from healthy controls, performed statistical analyses and revised the manuscript; S. S. - performed clinical assessment of patients, selected patients and revised the manuscript; S.R. - wrote and tested both algorithms in Java for calculating HFD and SampEn and revised the manuscript; M.Lj. – critically interpreted the data and wrote the manuscript; D.S.- critically interpreted the data and wrote the manuscript.

**Competing Financial Interests Statement:** The authors declare no competing financial interests.